# Reconstruction of Partial Dissimilarity Matrices for Cognitive Neuroscience

Denise Moerel [1,^], Tijl Grootswagers [1,2]

1| The MARCS Institute for Brain, Behaviour and Development, Western Sydney University, Sydney, Australia
2| School of Computer, Data and Mathematical Sciences, Western Sydney University, Sydney, Australia

^ corresponding author: d.moerel@westernsydney.edu.au

**Abstract**

In cognitive neuroscience research, Representational Dissimilarity Matrices (RDMs) are often incomplete because pairwise similarity judgments cannot always be exhaustively collected as the number of pairs rapidly increases with the number of conditions. Existing methods to fill these missing values, such as deep neural network imputation, are powerful but computationally demanding and relatively opaque. We introduce a simple algorithm based on geometric inference that fills missing dissimilarity matrix entries using known distances. We use tests on publicly available empirical cognitive neuroscience datasets, as well as simulations, to demonstrate the method's effectiveness and robustness across varying sparsity and matrix sizes. We have made this geometric reconstruction algorithm, implemented in Python and MATLAB, publicly available. This method provides a fast and accurate solution for completing partial dissimilarity matrices in the cognitive neurosciences.

# 1 Introduction

Representational Similarity Analysis (RSA; Kriegeskorte et al., 2008) is a popular approach used in cognitive neuroscience to compare patterns of neural or behavioural responses across stimuli (Cichy et al., 2014; Cichy and Oliva, 2020; Contini et al., 2017; Grootswagers et al., 2024, 2019, 2017; Koenig-Robert et al., 2024; Moerel et al., 2024b, 2024a; Robinson et al., 2023). The key input is a Representational Dissimilarity Matrix (RDM), a symmetric matrix summarising pairwise distances or dissimilarities.

In experimental settings, collecting all pairwise similarity judgments for large numbers of stimuli is often impractical or impossible. For instance, in behavioural triplet tasks, where participants are presented with three items and are asked to pick the one that is the most different from the other two (Hebart et al., 2020), the number of trials needed for a complete estimate scales quickly. Alternative tasks have been used that do not require iterating through every individual pair, such as inverse multidimensional scaling (Kriegeskorte and Mur, 2012), but these tasks scale quickly in completion time. In many cases it is not practically feasible to collect enough trials resulting in sparse dissimilarity matrices with many missing entries. Sparse dissimilarity matrices present challenges for subsequent model correlations in Representational Similarity Analysis and in estimating embedding spaces, as many algorithms for those methods cannot handle incomplete dissimilarity matrices or result in less statistical power.

Recent work in cognitive neuroscience has started to address this challenge by modelling similarity matrices for natural objects using predictive or deep learning approaches (Fu et al., 2023; Gifford et al., 2022; Hebart et al., 2020; Kaniuth et al., 2025), showing that dissimilarity matrices can be inferred from limited data. These methods involve for example using deep neural networks to efficiently predict perceived similarity from input images. However, these methods still require relatively large training datasets, and sometimes need to be trained on the input stimuli, which means they are not widely applicable.

Here, we propose a simple geometric reconstruction method that completes missing dissimilarity matrix entries using Euclidean geometry based on known distances. We empirically assess reconstruction quality using both simulated and real similarity data and compare performance to related distance matrix completion approaches. Our method provides a fast and transparent alternative tailored for sparse Representational Dissimilarity Matrices. We have implemented this algorithm in Python and MATLAB, in an easy-to-use function which only requires the matrix with missing values. The code is publicly available and can be found here: https://github.com/Tijl/completeRDM.

# 2 Methods

## 2.1 Geometric reconstruction algorithm

We start with an incomplete Representational Dissimilarity Matrix represented as a square symmetric distance matrix where several pairwise distances are missing. For each missing entry (e.g., *nan* or *none*) representing the distance between stimulus i and j, we find the set of entries k where the distances from i to k and j to k are known.

Assuming Euclidean geometry and triangle inequality, we estimate the missing distance using either as the square root of the sum of squares or the difference of squares of these known distances. Specifically, for each valid k, two estimates are calculated: $(d(i,k)^2 + d(j,k)^2)^{0.5}$ and $(|d(i,k)^2 - d(j,k)^2|)^{0.5}$. The median of all these estimates for all valid k is used as the predicted distance. The algorithm operates by iteratively filling missing entries using known distances to other points. If no valid estimates exist for a missing value on the current pass, the algorithm skips that value and tries again in the next iteration after more missing entries have been filled in. The process completes when no more entries can be filled. If there exists at least one path between any two entries where all distances are known (i.e., no disconnected components), the algorithm is guaranteed to fill the entire matrix. If there are isolated clusters, the algorithm fills only within connected components.

## 2.2 Evaluating performance

To evaluate the performance of the algorithm, we used publicly available data from four previously published works. We opted for behavioural similarity datasets as this is the most likely case where collecting all pairwise data is not feasible for larger dissimilarity matrices. The first dataset has behavioural similarity ratings for 26 images (Bracci et al., 2019). The second used a triplet task for 36 abstract object concepts (Robinson et al., 2025). The third dataset were pairwise similarities obtained using an inverse multidimensional scaling approach for 92 objects (Mur et al., 2013). The fourth dataset used a triplet task for 256 simple grating stimuli (Grootswagers et al., 2024). These four studies were selected as their data were publicly available, the dissimilarity matrices were complete and contained no missing entries, and they presented variation in the type of stimulus, type of task, and total amount of data. Using 1000 random iterations, we systematically deleted an increasing percentage (1-80%) of pairwise distances and used the algorithm to reconstruct the missing values. Reconstruction accuracy was measured as the correlation between the reconstructed data and the original data. We calculated the accuracy separately for just the reconstructed missing values, as well as the entire reconstructed matrix.

We also assessed reconstruction performance using a more systematic simulation approach, where we calculated full distance matrices on randomly placed points. We generated data of four sizes (32 × 32, 64 × 64, 128 × 128, and 256 × 256). Similar to the empirical data approach, we systematically deleted a percentage (1-80%) of pairwise distances and used the algorithm to reconstruct the missing values. We repeated this for 1000 iterations and calculated reconstruction accuracies as before through the correlation between the reconstructed and ground truth data.

Finally, we compared the performance of the geometric approach to two alternative methods. The first alternative is a graph-based method, where the known distances in the dissimilarity matrix are the edges between nodes in a graph. We then use a variant of the Floyd-Warshall algorithm to reconstruct missing edges, by iteratively calculating the shortest path between two nodes with missing edges. The second alternative is to use multi-dimensional scaling on the partial dissimilarity matrix to estimate an embedding, from which a full dissimilarity matrix is then inferred. We compared reconstruction performance of all three algorithms on the four empirical datasets using 100 simulations at 1-80% percentage missing pairwise distances.

# 3 Results

We evaluated the geometric reconstruction method using simulated data, empirical dissimilarity matrices from previous works, and by comparing the geometric approach to other distance matrix completion methods.

## 3.1 Empirical Data

We applied the reconstruction method to four Representational Dissimilarity Matrices from previously published studies (Bracci et al., 2019; Grootswagers et al., 2024; Mur et al., 2013; Robinson et al., 2025). We then artificially masked entries to simulate missing data patterns typical in behavioural similarity tasks and compared the accuracy of the reconstructed matrices with the originals using Pearson correlations. We tested matrices of sizes 27 × 27, 36 × 36, 92 × 92, and 256 × 256, with varying proportions of missing distances from 1% to 80%. For each size and proportion combination, random distances were removed, and the geometric reconstruction algorithm was applied. This was repeated 1000 times for each size and proportion combination to estimate a range of reconstruction accuracies. Reconstruction accuracy was assessed by correlating the estimated distances with the true distances on the missing entries. We did this for the reconstructed values only (Figure 1A) as well as the full dissimilarity matrix (Figure 1B). Example reconstructions are shown in Figure 1C. The results show that the reconstruction accuracy decreased systematically with increasing number of missing values, but consistently outperformed chance (inserting random values), even when up to 80% of the original data was missing. In addition, we observed that overall reconstruction accuracy was higher for larger dissimilarity matrices. Together, these results show that the reconstruction method successfully reconstructed the removed distances. This demonstrates practical utility for this reconstruction method, making it highly suitable for incomplete similarity matrices in cognitive neuroscience.

## 3.2 Simulated Data

It is possible that the inherent structure in the empirical dissimilarity matrices leads to underestimating reconstruction error rates. To address this, we generated synthetic datasets without inherent structure by sampling vectors from a multivariate Gaussian distribution to create an embedding of stimuli in a Euclidean space. From these embedding vectors, we computed the full Euclidean distance matrices as ground truth. We tested matrices of sizes 32 × 32, 64 × 64, 128 × 128, and 256 × 256, with varying proportions of missing distances from 1% to 80%. Following the analysis on the real data, we masked random distances as missing for each size and proportion combination and used the recursive imputation algorithm to reconstruct the full dissimilarity matrix. We repeated this 1000 times for each size and proportion combination. Figure 2 shows that for these simulated data without any inherent underlying embedding structure, we also obtained high reconstruction accuracies across all dissimilarity matrix sizes, and that reconstruction accuracy decreased with more missing data. Consistent with the findings on the real data, we also observed that the decrease as a function of missing values was stronger for smaller matrix sizes.

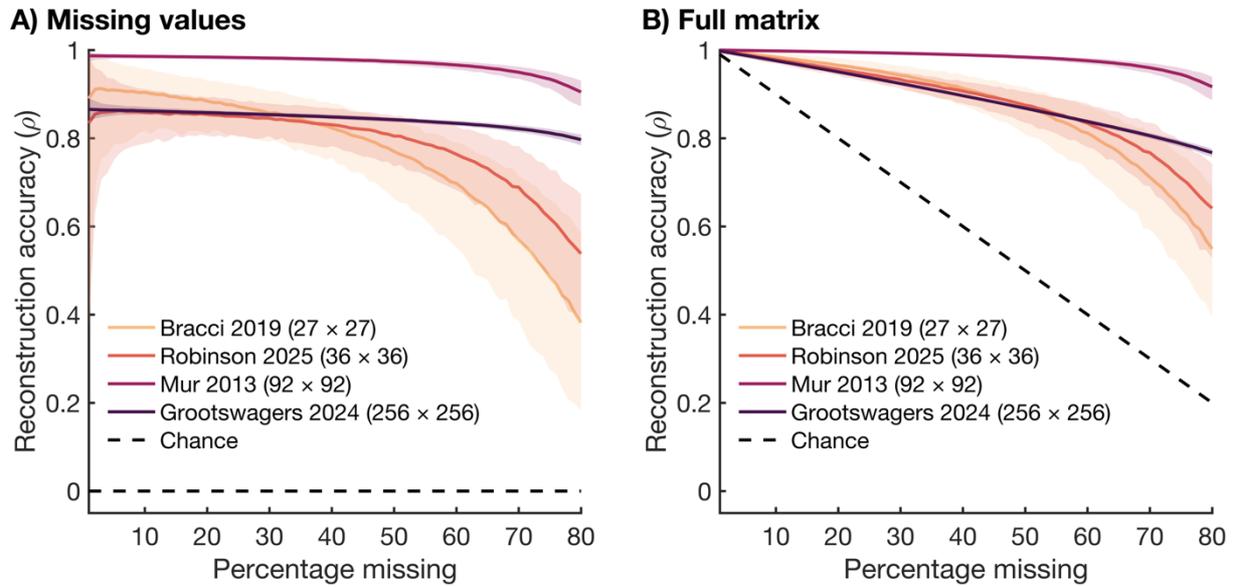
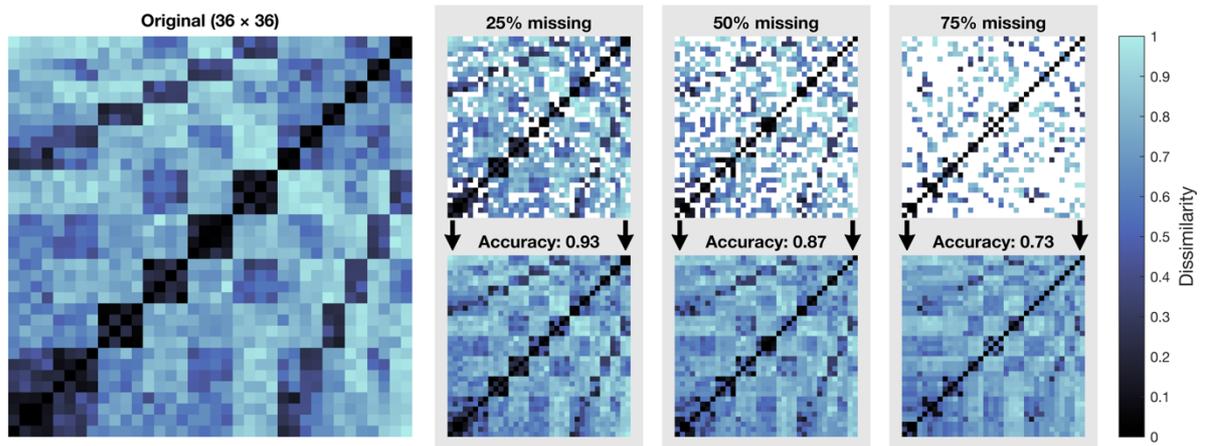

**Figure 1. Reconstruction accuracies for the empirical data.** A) The reconstruction accuracies for the missing values only, across varying proportions of missing distances. The reconstruction accuracy was calculated as the correlation between the original and reconstructed distances. We used 4 publicly available datasets, shown in different colours. The shaded area around the plot lines shows the 2.5 to 97.5 percentile. The black dashed line indicates theoretical chance, assuming missing values are substituted for random values. B) The reconstruction accuracies for the full dissimilarity matrices, across varying proportions of missing distances. All plotting conventions are the same as Figure 1A. C) Example reconstructions, based on a 36 × 36 dissimilarity matrix (Robinson et al., 2025). The left panel shows the original full dissimilarity matrix. Lighter colours indicate greater dissimilarity. The remaining panels show, from left to right, the dissimilarity matrices with missing values indicated in white (top), and the reconstructed dissimilarity matrix (bottom) for 25%, 50%, and 75% missing values.

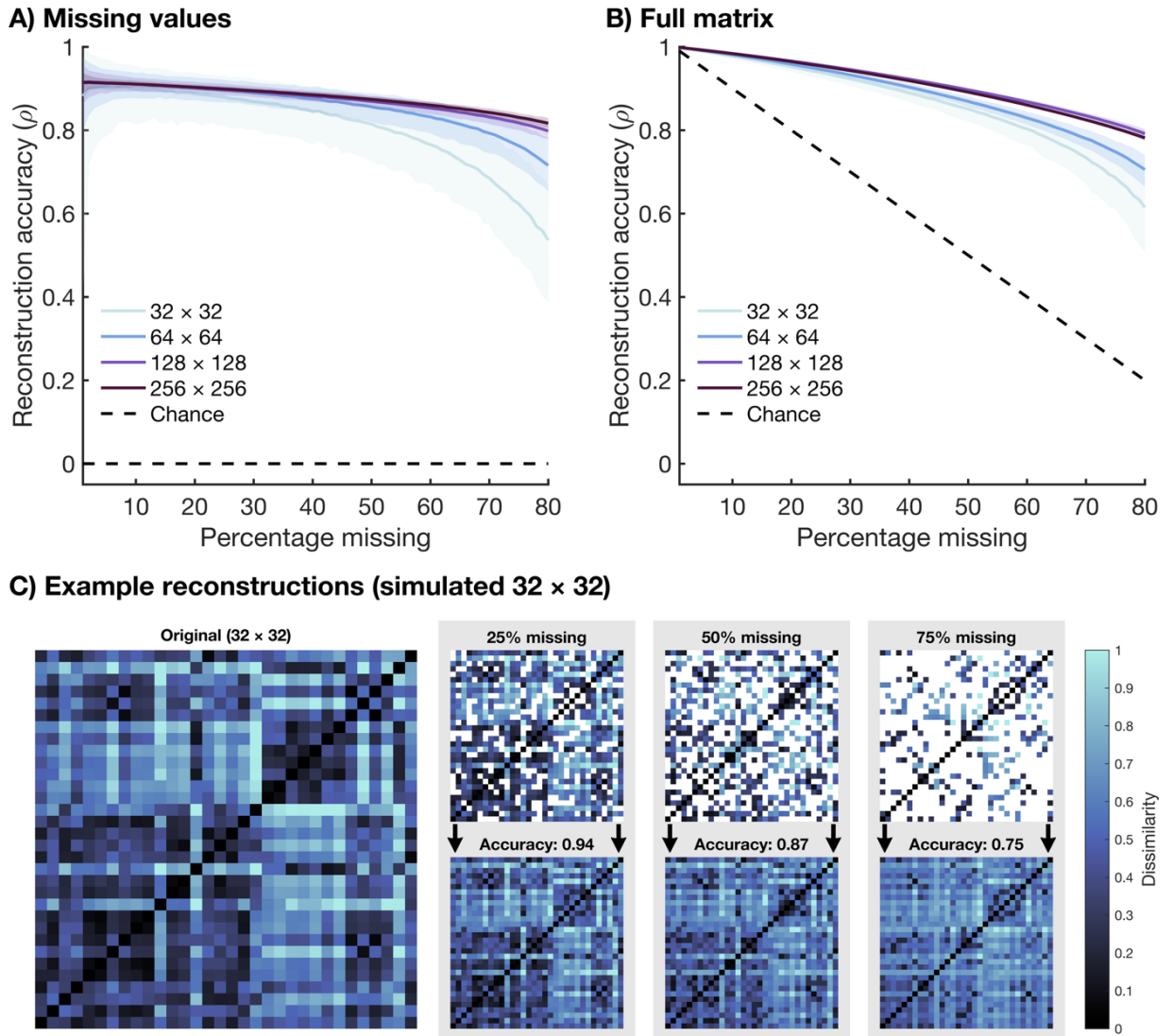

**Figure 2. Reconstruction accuracies for the simulated data.** A) The reconstruction accuracies for the missing values only, across varying proportions of missing distances. The reconstruction accuracy was calculated as the correlation between the original and reconstructed distances. We used 4 different sized dissimilarity matrices, shown in different colours. The shaded area around the plot lines shows the 2.5 to 97.5 percentile of the 1000 simulations. The black dashed line indicates theoretical chance, assuming missing values are substituted for random values. B) The reconstruction accuracies for the full dissimilarity matrices, across varying proportions of missing distances. All plotting conventions are the same as Figure 2A. C) Example reconstructions, based on a 32 × 32 dissimilarity matrix. The left panel shows the original full dissimilarity matrix, with lighter colours indicating greater dissimilarity. The remaining panels show, from left to right, the dissimilarity matrices with missing values indicated in white (top), and the reconstructed dissimilarity matrix (bottom) for 25%, 50%, and 75% missing values.

### 3.3 Comparison with alternative reconstruction methods

Finally, we compared the geometric reconstruction method to two other methods for completing distance matrices. We used a graph-based method, and a method based on multidimensional scaling. On the four empirical datasets, we calculated reconstruction accuracies between 1% to 80% missing values. Figure 3 shows that the geometric model overall outperforms the two other methods, with the graph-based method losing accuracy quickly with increasing missing values, for all except the largest matrix (256 × 256). The method using multidimensional scaling showed lower reconstruction accuracies, but also much larger variance compared to the other methods. Together, these results show that the geometric reconstruction method outperforms the alternative methods.

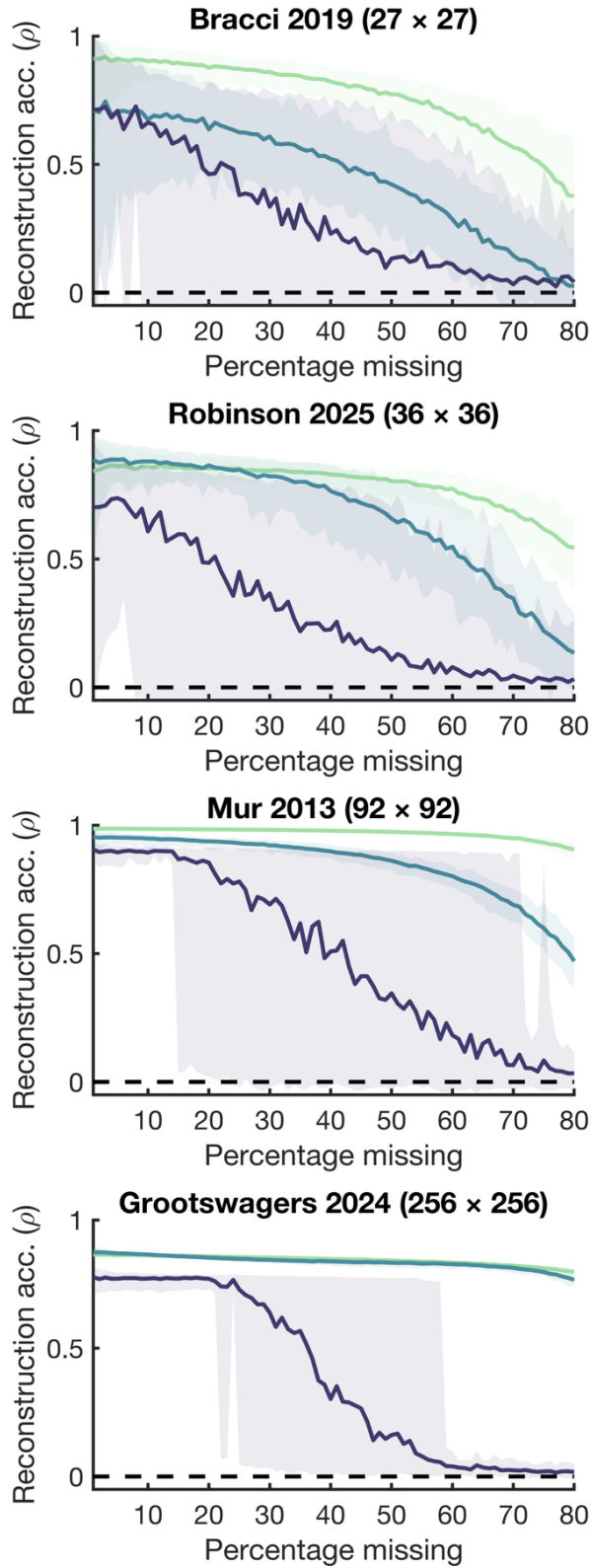
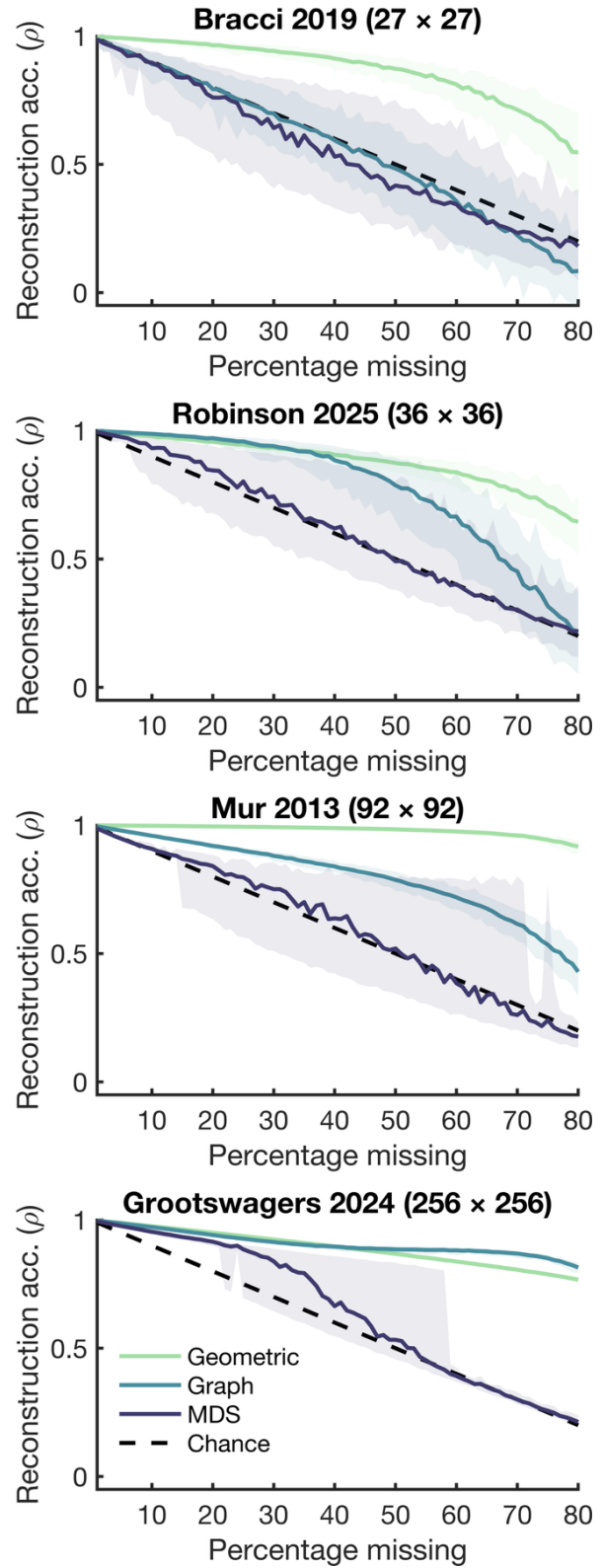

**Figure 3. Comparing alternative reconstruction approaches.** The reconstruction accuracy is calculated as the correlation between the original and reconstructed distances. Each row represents the reconstruction results on a different empirical dataset. The different lines represent different methods; geometric, graph, and multidimensional scaling (MDS). The shaded area around the plot lines shows the 2.5 to 97.5 percentile of the 100 simulations. The black dashed line indicates theoretical chance, assuming missing values are substituted for random values. A) the reconstruction accuracies on missing values only. B) the reconstruction accuracies on the full dissimilarity matrix.

## 4 Discussion

Our recursive geometric algorithm offers a simple, transparent way to fill missing values in sparse dissimilarity matrices using Euclidean geometry. It requires no complex training or prior models, relying solely on known distances and Euclidean geometry. The algorithm runs in polynomial time, with a worst-case time complexity of $O(n^3)$ when the number of missing values is maximal, making it computationally tractable for typical dissimilarity matrix sizes in cognitive neuroscience. Simulations and empirical results showed that the overall reconstruction accuracy was high, and did not decrease as much with increasing numbers of missing values compared to alternative methods. Strikingly, even with as little as 20% of values remaining, the geometric reconstruction approach resulted in a correlation of 0.8 with the original data. An exception was the 92 × 92 dataset (Mur et al., 2013), for which performance remained close to ceiling regardless of the missing values. One possible explanation is that this dataset contains an extremely strong clustering, a result that has been observed repeatedly on this benchmark stimulus set (cf. Grootswagers and Robinson, 2021).

The main assumption underlying the geometric reconstruction approach is that the representational space approximately follows Euclidean geometry, meaning entries in the dissimilarity matrix can be embedded in a multidimensional flat metric space where distances obey triangle inequalities. According to our results on empirical data, this assumption holds reasonably well for cognitive (neuro)science datasets. The reconstruction method may break down if the true space is curved or non-metric. Future work could explore extending this approach to accommodate non-Euclidean geometries. Despite this potential limitation, our simulations with synthetic data and empirical dissimilarity matrices show robust and accurate reconstruction across varying matrix sizes and missing data levels. This demonstrates the method's practical value for tasks with partial similarity information, where obtaining pairwise similarities for all possible pairs of stimuli is infeasible.

In conclusion, here we described and evaluated a geometric method as an interpretable, lightweight solution for reconstructing sparse dissimilarity data, supporting researchers in obtaining complete dissimilarity matrices from incomplete measurements.

## 5 Data and Code availability

All data used in this paper is publicly available (Bracci et al., 2019; Grootswagers et al., 2024; Mur et al., 2013; Robinson et al., 2025). The geometric reconstruction algorithm, implemented in Python and MATLAB, is publicly available here: https://github.com/Tijl/completeRDM. This repository also contains all analysis and plotting code used in this paper.

# 6  Funding

TG is supported by the Australian Research Council (DE230100380).